\documentclass[showpacs,preprintnumbers,amsmath,amssymb,twocolumn,superscriptaddress,prl]{revtex4}
\usepackage{graphicx}
\usepackage{amsfonts}
\usepackage{amsmath, amsthm, amssymb}
\usepackage{dsfont}
\usepackage{times}
\usepackage{subfigure}

\newcommand{\bq}{\begin{equation}}
\newcommand{\eq}{\end{equation}}
\newcommand{\eps}{\varepsilon}

\newcommand{\etal}{\emph{et al.}}
\def\i{\mathrm{i}}

\begin{document}
\title[Pairing symmetry conversion by spin-active interfaces in superconducting junctions]{
Pairing symmetry conversion by
spin-active interfaces in superconducting junctions}

\author{Jacob Linder}
\affiliation{Department of Physics, Norwegian University of
Science and Technology, N-7491 Trondheim, Norway}
\author{Takehito Yokoyama}
\affiliation{Department of Applied Physics, Nagoya University, Nagoya, 464-8603, Japan}
\author{Asle Sudb{\o}}
\affiliation{Department of Physics, Norwegian University of
Science and Technology, N-7491 Trondheim, Norway}
\author{Matthias Eschrig}
\affiliation{Institut f{\"u}r Theoretische Festk{\"o}rperphysik and DFG-Center for Functional Nanostructures,\\
Universit{\"a}t Karlsruhe, D-76128 Karlsruhe, Germany}

\date{Received \today}
\begin{abstract}
\noindent 
We study the proximity-induced superconducting correlations in a 
normal metal connected to a superconductor when the interface between them 
is spin-active and the normal metal is ballistic or diffusive. 
Remarkably, for any interface spin polarization there is a critical interface
resistance, above which
the conventional even-frequency proximity component vanishes completely 
at the chemical potential,
while the odd-frequency component remains finite.
We propose a way to unambiguously observe 
the odd-frequency component.
\end{abstract}

\maketitle
\par
Superconductivity and superfluidity are hallmarks of the wave-like character 
of matter, and manifest themselves in vastly different systems, from ultracold 
dilute gases via cold metals and fluids, to extremely dense
protonic and neutronic matter. In all these 
contexts, the symmetry of the order parameter is of profound importance. Over the 
last decades, the possibility of superconducting order parameters that change sign 
under a {\it time-coordinate} exchange of the two fermions comprising the Cooper-pair, 
has emerged in addition to the by now 
well studied varieties of orbital symmetries
\cite{bergeretrmp,buzdinrmp,tanakaPRL,EschrigLTP,Yokoyama}. 
This so-called odd-frequency superconductivity \cite{berezinskii} 
is distinct from the traditional even-frequency pairing in the 
Bardeen-Cooper-Schrieffer paradigm, and may be induced by proximity effects in hybrid 
structures of superconductors and magnets \cite{bergeretrmp}. 
\par 
In a broader context, proximity systems offer the possibility of 
controlling the physics of competing broken symmetries.  
The fundamental heterostructure for studying proximity induced superconductivity
is the superconductor/normal metal (S$\mid $N) bilayer, where the normal 
metal or the interface may have magnetic properties. 
Among possible triplet pair correlations, in the diffusive limit odd-frequency pairs
are favored \cite{volkovPRL}, whereas in 
ballistic hybrid systems both odd- and even-frequency amplitudes
compete \cite{tanakaPRL,EschrigLTP}.  As all known superconductors to 
date exhibit an even-frequency order parameter, 
the observation of proximity induced effects that 
are particular to odd-frequency pairing would be of utmost interest.
\par
There are two major difficulties associated with the detection of the 
odd-frequency state in superconductor/ferromagnet (S$\mid$F) bilayers.
One is the usually short penetration depth into the ferromagnetic region, limited 
by the magnetic coherence length $\xi_F$, much less than the 
superconducting coherence length $\xi_S$ \cite{bergeretrmp}. 
Another problem is that odd-frequency pairs are only
well defined when even-frequency correlations vanish in the ferromagnet.
Clear-cut signatures of the former are therefore only accessible in a 
limited parameter regime \cite{linder}. 
\par
The majority of work on superconducting proximity-structures so far has been 
restricted to the diffusive limit and spin-inactive interfaces \cite{kupluk}. 
For a non-magnetic bilayer, a minigap appears 
in the density of states of the normal metal. It scales with the Thouless 
energy of the normal layer and with the transmission probability of the interface. 
Such minigap structures are readily accessible experimentally \cite{sueur_prl_08}.
For a spin-active interface, the transmission properties of spin-$\uparrow$ and 
spin-$\downarrow$ electrons into a metal are different, and this 
gives rise to both spin-dependent conductivities and spin-dependent phase shifts 
at the interface \cite{Millis,hh,Brataas,eschrig,audrey}. In this Letter we show
that a spin-active interface in a S$\mid $N bilayer produces clear signatures of
purely odd-frequency triplet pairing amplitudes that can be tested experimentally.
\par
\begin{figure}[b!]
\centering
\resizebox{0.35\textwidth}{!}{
\includegraphics{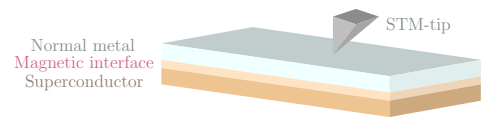}}
\caption{(Color online) Proposed experimental setup for observation of the odd-frequency 
component in a diffusive normal metal layer$\mid$superconductor junction.}
\label{fig:model} 
\end{figure}
We consider the system shown in Fig. \ref{fig:model}. The superconductor is conventional 
(even-frequency $s$-wave) while the interface is magnetic.
We find that there is a dramatic change in the nature of proximity 
correlations when the spin-dependent phase shifts 
exceed the tunneling probability of the interface. The spin-active interface 
in an S$\mid$N bilayer causes
the even-frequency correlations to vanish at zero excitation energy,
while odd-frequency correlations appear.  At the same time, the 
minigap, one of the hallmarks of the conventional proximity effect, 
is replaced by a low-energy band with enhanced density of states.
We focus on the density of states (DOS) in the normal region, which
can be probed by tunneling experiments. 
Our findings  suggest that  it should be possible to detect 
the odd-frequency amplitude without any interfering effects of even-frequency 
correlations.  Since the exchange field is absent in the normal metal, 
this resolves the two main difficulties associated with the 
experimental detection of odd-frequency correlations mentioned above.
\par
We adopt the quasiclassical theory of superconductivity \cite{quasiclassical}, 
where information about the physical properties of the system is embedded in 
the Green's function. For equilibrium situations, it suffices to consider 
the retarded Green's function, $\hat{g}$, that is parameterized conveniently 
in the normal (N) region by a parameter $\theta_\sigma$, allowing for both 
singlet and triplet correlations \cite{linder}. In the superconducting (S) 
region, we employ the bulk solution
$\hat{g}_S = c\cdot\underline{\tau_3}\otimes\underline{\sigma_0} + 
s \cdot \underline{\tau_1}\otimes(\i\underline{\sigma_2})$, with
$c=\cosh(\theta)$, $s=\sinh(\theta)$, $\theta=\text{atanh}(\Delta/\varepsilon)$,
$\underline{\tau_i}$ and $\underline{\sigma_i}$ being Pauli matrices in 
particle-hole and spin space, respectively.
\par
We use the formalism described in Ref. \cite{linder}, and consider first 
the diffusive limit.  Then, the orbital symmetry for all 
proximity amplitudes is reduced to $s$-wave and hence the singlet component 
always has an even-frequency symmetry while the triplet component has an 
odd-frequency symmetry.  The Green's functions are subject to boundary conditions,
which in the tunneling limit  assume the following form at the S$\mid$N interface 
 \cite{hh,audrey}: 
$2\gamma d \hat{g}_N \partial_x \hat{g}_N = [\hat{g}_S,\hat{g}_N] + \i (G_\phi/G_T) [
\underline{\tau_0}\otimes\underline{\sigma_3} , \hat{g}_N],$
and at the outer interface read $\partial_x \hat{g}_N =\hat 0 $.
Here, $\gamma = R_B/R_N$ where $R_B$ $(R_N)$ is the resistance of the barrier 
(normal region), and $d$ is the width of the normal region, while $G_T$ is the 
junction conductance in the normal-state. The boundary condition above contains 
an additional term $G_\phi$ compared to the usual non-magnetic boundary conditions 
in Ref. \cite{kupluk}.  This term is due to spin-dependent phase shifts of 
quasiparticles being reflected at the interface.  $G_\phi$ may be non-zero even 
if the transmission $G_T \to 0$, corresponding to a ferromagnetic 
insulator \cite{hh}. We define the superconducting coherence length 
$\xi_S = \sqrt{D/\Delta}$ and Thouless energy $\varepsilon_\text{Th}=D/d^2$, 
where $D$ is the diffusion constant, and assume that the inelastic scattering 
length, $l_\text{in}$, is sufficiently large, such that $d\ll l_\text{in}$.
\par
The Usadel equation \cite{usadel} reads 
$D\partial_x^2\theta_\sigma + 2\i\varepsilon\sinh\theta_\sigma= 0$,
with boundary condition $\gamma d \partial_x\theta_\sigma = 
(c s_\sigma - \sigma sc_\sigma) + \i\sigma s_\sigma 
\frac{G_\phi}{G_T}$
at $x=0$ and $\partial_x\theta_\sigma = 0$ at $x=d$. Here, 
$c_\sigma=\cosh(\theta_\sigma)$ and $s_\sigma=\sinh(\theta_\sigma)$. 
At zero energy, we find that the pairing amplitudes are either purely 
(odd-frequency) triplet,
\begin{align}\label{eq:full1}
f_s(0) = 0,\; f_t(0) = \frac{G_T \cdot \text{sgn}(G_\phi )}{\sqrt{G_\phi^2-G_T^2}}\;\; 
\text{ for } \frac{|G_\phi|}{G_T} > 1,
\end{align}
or purely (even-frequency) singlet
\begin{align}\label{eq:full2}
f_s(0) = \frac{\i\cdot G_T}{\sqrt{G_T^2-G_\phi^2}},\; f_t(0) = 0\;\; 
\text{ for } \frac{|G_\phi|}{G_T} < 1.
\end{align}
Thus, the presence of $G_\phi$ induces an odd-frequency component in the normal layer. 
The remarkable aspect of Eqs. (\ref{eq:full1}) and (\ref{eq:full2}) is that they 
are valid for any value of the width $d$ below the inelastic scattering length, and for
any interface parameter $\gamma$. Thus, the vanishing of the singlet component 
is a robust feature in S$\mid$N structures with spin-active interfaces, as long 
as $|G_\phi|/G_T>1$. Without loss of generality, we focus on positive 
values of $G_\phi$ from now on. 
\begin{figure}[t!]
\centering
\resizebox{0.5\textwidth}{!}{
\includegraphics{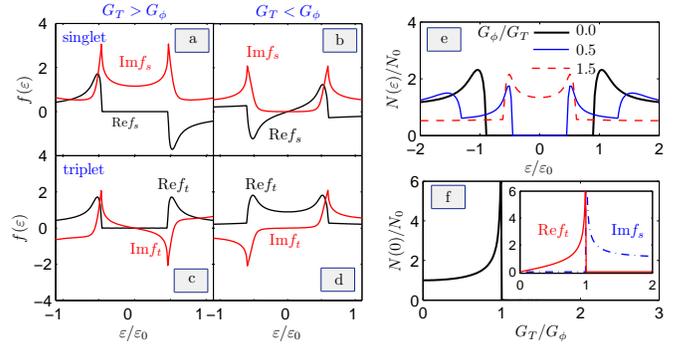}}
\caption{(Color online) The singlet and triplet proximity amplitudes induced in the 
normal metal are shown for $G_\phi/G_T<1$ [in a) and c)] and $G_\phi/G_T>1$ 
[in b) and d)]. In e), we plot the energy-resolved DOS for several values 
of $G_\phi/G_T$. Finally, f) shows the zero-energy DOS as a function of $G_T/G_\phi$, 
with the proximity amplitudes shown in the inset.}
\label{fig:anomalous_diffusive} 
\end{figure}
The DOS is given as $N(\varepsilon)/N_0 = \sum_\sigma \text{Re}\{c_\sigma\}/2$, 
yielding $N(0)/N_0 = \text{Re}\{G_\phi/\sqrt{G_\phi^2 -G_T^2}\}$. At zero-energy, 
the DOS thus vanishes as long as $G_\phi/G_T < 1$, which means that the usual 
minigap in S$\mid$N structures survives in this regime. However, the zero-energy DOS is 
enhanced for $G_\phi/G_T > 1$ since the singlet component vanishes there. 
\par
The full energy-dependence of the DOS may only be obtained numerically. 
To model a realistic experimental setup, we fix $\gamma=10$ and $d/\xi_S=1.0$, 
although our qualitative results are independent of these particular choices. 
As a measure of the relevant energy scale, we define 
$\varepsilon_0 = \varepsilon_\text{Th}/(2\gamma)$. 
The results are shown in Fig. \ref{fig:anomalous_diffusive} 
to investigate the effect of the spin-dependent phase shifts. 
The low-energy DOS is strongly enhanced due to the 
odd-frequency amplitude when $G_\phi/G_T>1$ ($G_\phi/G_T=1.5$ in the figure). 
Conversely, the DOS develops a minigap around $\varepsilon=0$ when 
$G_\phi/G_T<1$ ($G_\phi/G_T=0.5$ in the figure).  The ratio $G_\phi/G_T$ depends 
on the microscopic barrier properties \cite{audrey}. In the tunneling limit, 
one finds that $G_\phi$ can be considerably larger than $G_T$. 
\par
We suggest the following qualitative explanation for the mechanism behind the 
separation between even- and odd-frequency correlations. The superconductor 
induces a minigap $ \propto G_T$ in the normal metal, while the spin-active 
barrier induces an effective exchange field $ \propto G_\phi$. The 
situation in the normal metal then resembles that of a thin-film conventional 
superconductor in the presence of an in-plane external magnetic field 
\cite{meservey}, with the role of the gap and field played by $G_T$ and 
$G_\phi$, respectively. In that case, it is known that superconductivity is 
destroyed above the Clogston-Chandrasekhar limit \cite{clogston}, as the 
spin-singlet Cooper-pairs break up.  In the present case, we 
observe coexistence of the exchange field and spin-singlet even-frequency 
superconductivity as long as $G_\phi$ is below the critical value of $G_\phi=G_T$. 
However, for $G_\phi>G_T$  spin-singlet pairing is no longer possible at 
the chemical potential. It is then replaced by spin-triplet pairing, which must 
be odd in frequency due to the isotropization of the gap in the diffusive limit. Thus, 
there is a natural separation between even-frequency and odd-frequency pairing 
in the normal metal at a critical value of the effective exchange field $G_\phi$.
\par
The same effect occurs in the ballistic limit, as we now show. In this case, we can 
obtain the retarded Green's function using the formalism described in 
Refs. \cite{eschrig,fogelstrom}.  The Eilenberger equation in the normal region reads 
$i v_{Fx} \partial_x \hat g + [\eps \underline{\tau_3}\otimes\underline{\sigma_0} 
,\hat g] =\hat 0$.  For the boundary conditions, we use a scattering matrix 
describing the magnetic interface between the superconductor and the normal metal,
\begin{equation}
\hat S= \left(
\begin{array}{cc}
\underline{r}_S \cdot \exp \left(\frac{i}{2}\vartheta_S\underline{\sigma_3}\right) 
& \underline{t}_{SN}\cdot \exp\left(\frac{i}{2}\vartheta_{SN}\underline{\sigma_3}\right)\\
\underline{t}_{NS}\cdot \exp \left(\frac{i}{2}\vartheta_{NS}\underline{\sigma_3}\right) 
& -\underline{r}_N\cdot \exp \left(\frac{i}{2}\vartheta_N\underline{\sigma_3}\right)
\end{array} \right) ,
\label{SM}
\end{equation}
with real reflection and transmission spin matrices 
$\underline{r}_S$, $\underline{r}_N$, $\underline{t}_{SN}$, and $\underline{t}_{NS}$.
The spin mixing angles $\vartheta_S$, $\vartheta_N$, $\vartheta_{SN}$, 
and $\vartheta_{NS}$ 
describe spin dependent scattering phases \cite{Millis}.
Neglecting spin flip scattering,
the transmission and reflection amplitudes are diagonal in spin space,
and the relations $\underline{r}_S=\underline{r}_N\equiv \mbox{diag}\left[ r_\uparrow,r_\downarrow \right]$, $\underline{t}_{NS}=\underline{t}_{SN}\equiv \mbox{diag}\left[t_\uparrow, t_\downarrow \right] $, 
$r_\uparrow^2+t_\uparrow^2=r_\downarrow^2+t_\downarrow^2=1$,
$\vartheta_{NS}+\vartheta_{SN}
=\vartheta_S+\vartheta_N$ follow from the unitarity of $\hat S$.
Possible scalar phases are omitted in Eq. \eqref{SM}, as they play no role in the final results.
\par
We next concentrate on subgap energies. The anomalous amplitudes  can be 
decomposed into singlet and triplet components, 
$f=(f_s+f_t\, \underline{\sigma_3}) (i\underline{\sigma_2})$. Defining $f_\sigma = 
(f_s + \sigma f_t)/2$, we obtain
on the top of the normal overlayer ($x=d$) 
$f_\sigma (\eps )= -\mbox{sgn} (\alpha_\sigma )
t_\uparrow t_\downarrow /\sqrt{\alpha_\sigma^2-(t_\uparrow t_\downarrow )^2}$
with $\alpha_\sigma = \sin \left( 2\eps d/v_{Fx} + \sigma \vartheta_+ \right)
+r_\uparrow r_\downarrow \sin \left( 2\eps d/v_{Fx} + \sigma \vartheta_- \right)$.
Here, $\vartheta_\pm = \frac{1}{2}(\vartheta_N \pm \vartheta_S)\pm 
\arcsin (\eps /\Delta )$, and $\eps  $ has to be supplemented by an infinitesimally 
small positive imaginary part. The interface parameters and the
Fermi velocity component in $x$-direction, $v_{Fx}=v_F\cos \psi$, depend on the impact
angle $\psi$. The relevant energy scale in the problem is the ballistic 
Thouless energy, $\eps_{Th}=v_{F}/2d$.  For zero spin mixing angles we recover 
the known DOS for a normal state overlayer on a singlet superconductor.
The DOS is non-zero only for $|\alpha |>t_\uparrow t_\downarrow $, which
for sufficiently large impact angle always is fulfilled.
Clearly, the most interesting regime concerns $\eps /\eps_{Th} \sim
|\vartheta_\pm |\sim t_\uparrow t_\downarrow $.
\par
In the tunneling limit, for small excitation energies $\eps /\eps_{Th}\ll 1$ and small
spin mixing angles $\vartheta_\pm $ we obtain $\alpha_\sigma = (4\eps d/v_{Fx} + 
\sigma \vartheta_N )$. In this case, due to $\vartheta_++\vartheta_- = \vartheta_N$,
only the spin mixing angle for reflection at the normal side of the interface enters, and
acts as an effective exchange field $b=\vartheta_N v_{Fx}/4d $ on the quasiparticles. 
Especially interesting is the case $\eps =0$, for which all proximity amplitudes 
are even in momentum.  For $\eps=0 $ we obtain $\alpha_\sigma = \sigma \vartheta_N $,
and the pairing amplitudes are either purely (odd-frequency) triplet,
\begin{equation}
f_s(0)=0,\; f_t(0)= \frac{-t_\uparrow t_\downarrow \!\cdot \! \mbox{sgn} (\vartheta_N) }{
\sqrt{\vartheta_N^2-(t_\uparrow t_\downarrow )^2}}\;\; \text{ for } 
\frac{|\vartheta_N|}{t_\uparrow t_\downarrow} > 1,
\end{equation}
or purely (even-frequency) singlet
\begin{equation}
f_s(0)= \frac{ \i \cdot t_\uparrow t_\downarrow }{\sqrt{(t_\uparrow t_\downarrow )^2-\vartheta_N^2}}, \; f_t(0)=0 
\;\; \text{ for } 
\frac{|\vartheta_N|}{t_\uparrow t_\downarrow } < 1 .
\end{equation}
Comparing with the results for the diffusive case, we find
that $G_\phi/G_T$ corresponds to $-\vartheta_N/(t_\uparrow t_\downarrow)$.
\par
\begin{figure}[t!]
\centering
\resizebox{0.45\textwidth}{!}{
\includegraphics{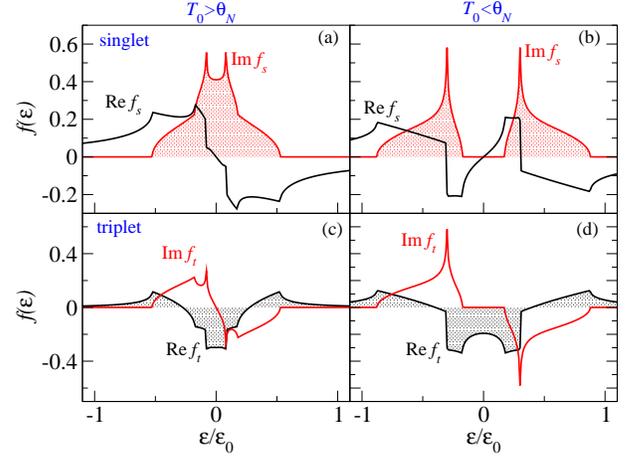}}
\caption{(Color online)
\label{fig:FigF} 
Momentum-averaged proximity amplitudes at the surface of the normal layer. 
Parameters: $d=v_F/\Delta$, $T_0 = 0.1$ (see text).  (a) and (c):
$\vartheta_N=\vartheta_S=0.05 <T_0$; (b) and (d):
$\vartheta_N=\vartheta_S=0.15 >T_0$.  Energy units are 
$\varepsilon_0= T_0 \, \varepsilon_{Th} $.  Even frequency singlet components 
are shown in (a-b), odd frequency triplet components in (c-d).}
\end{figure}
\begin{figure}[t!]
\centering
\resizebox{0.48\textwidth}{!}{
\includegraphics{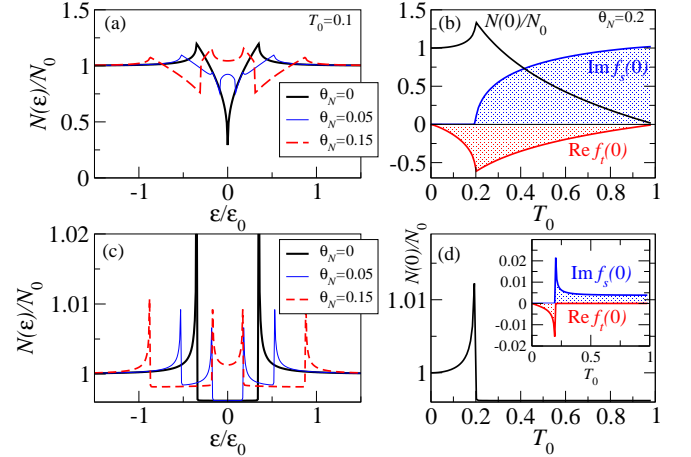}}
\caption{(Color online)
\label{fig:DOS_ball} 
(a) DOS as function of energy at the top of the normal layer for fixed transmission 
probability $T_0=0.1$, and various values of $\vartheta_N=\vartheta_S$. 
Remaining parameters are as in Fig. \ref{fig:FigF}. 
(b) DOS and proximity amplitudes at $\varepsilon=0$
for $\vartheta_N=\vartheta_S=0.2$ as function of $T_0$. 
In (c) and (d) we show the results corresponding to (a) and (b) 
when assuming an (abrupt) tunneling cone with opening angle of 10 degree.  }
\end{figure}
\par
In Fig. \ref{fig:FigF}, we show results for the proximity amplitudes in the 
ballistic limit, and focus on positive values of $\vartheta_N$ without 
loss of generality. 
A systematic expansion of all terms in the tunneling probability shows that
in the tunneling limit the spin dependence of the
transmission probabilities can be neglected, and only that of the phase shifts
needs to be kept. Thus, we assume $t_\uparrow=t_\downarrow =t$. We model the 
dependence on the impact angle $\psi $ as $t(\mu )=(t_0)^{\frac{1}{\mu} }$, 
$\mu=\cos \psi$, and assume for simplicity spin mixing angles independent of $\mu $. 
The tunneling probability for normal impact is $T_0=t_0^2$. 
In the case $T_0<\vartheta_N$ at small energies the odd frequency triplet 
amplitude dominates, and it is the only non-zero amplitude at $\eps=0$. On the 
other hand, for $T_0>\vartheta_N$ both singlet and triplet amplitudes 
contribute. This is due to the fact that for large impact angles the transmission 
probability $t(\mu)^2 $ drops below the value for the spin mixing angle $\vartheta_N $.
\par
We turn now to the DOS. The general expression, assuming the bulk solution in the 
superconductor, is $N(\eps)/N_0= \mbox{Re} \sum_{\sigma= \pm 1}
\int_0^1 |\alpha_\sigma |/\sqrt{\alpha_\sigma^2-(t_\uparrow t_\downarrow )^2} \; d\mu .$
In the tunneling limit, this simplifies again, and provided that 
$|\vartheta_N |> t_\uparrow t_\downarrow $ for all impact angles, the DOS at the 
Fermi level is enhanced above its normal state value,
$N(0)/N_0 =\int d\mu \; |\vartheta_N| / \sqrt{\vartheta_N^2-(t_\uparrow t_\downarrow)^2}$.
In Fig. \ref{fig:DOS_ball}, we show results for the DOS. In (a-b)
we assume the dependence on the impact angle as above, whereas in (c-d)
we allow tunnelling only in a narrow tunneling cone of 10 degrees. The DOS 
for the cases of dominating triplet amplitudes and dominating singlet amplitudes 
differ qualitatively.  In the case of a tunneling cone this difference 
is most drastic, and a comparison with the results above shows that it is very 
similar to the diffusive case.  In the right panels, where $\vartheta_N=\vartheta_S=0.2$, 
we demonstrate that for $T_0<0.2$ only the odd frequency triplet amplitude
is present at the chemical potential, while the singlet amplitude is
zero. The corresponding zero-bias DOS is enhanced in this region,
whereas it is reduced in the region when singlet correlations are present at $\eps =0$.
\par
The simplest experimental manifestation of the odd-frequency component is 
a zero-energy peak in the DOS \cite{Asano,yokoyama07,Braude}. 
In S$\mid$F layers, where this phenomenon has been discussed previously, 
a clear zero-energy peak is unfortunately often masked by the simultaneous presence 
of singlet correlations $f_s$, which tend to suppress the DOS at low energies. 
This is not so in the system we consider, provided only
$T_0<|\vartheta_N|$ in the ballistic limit, or equivalently, $G_T<|G_\phi |$ 
in the diffusive limit. This is ideal for a direct observation of the odd-frequency 
component,  manifested as a zero-energy peak in the DOS. 
\par
The important factor, with regard to isolation of the odd-frequency correlations 
at zero energy is the interface. The even-frequency correlations vanish completely 
when the interface transmission is sufficiently low. The parameters $\vartheta_N$, 
or equivalently, $G_\phi$ can be increased by increasing the magnetic polarization 
of the barrier separating the superconducting and normal layers. By fabricating 
several samples with progressively increasing strength of magnetic moment of the 
barrier, one should be able to observe an abrupt crossover at the zero-energy DOS 
above a certain strength of the magnetic moment. Alternatively, one could alter 
the interface transmission by varying the thickness of the insulating region.
\par
In summary, we have investigated the proximity-effect in a superconductor/normal 
metal bilayer with spin-active interface.  We find that both in the ballistic and 
diffusive limits, the usual even-frequency correlations may vanish completely 
at zero energy, while odd-frequency correlations persist. This result is completely 
independent of the specific values for the layer thicknesses and barrier resistances, 
indicating that it is a robust and general feature of spin-active interfaces. 
Our findings suggest a way of obtaining unambiguous experimental identification of 
superconducting odd-frequency correlations.
\par
\textit{Acknowledgments.} A. Cottet and D. Huertas-Hernando are thanked for helpful communications. J.L. and A.S. 
were supported by the Norwegian Research Council Grants No. 158518/431 and No. 158547/431 (NANOMAT), and Grant No. 
167498/V30 (STORFORSK). T.Y. was supported by the JSPS.


\vspace{-0.5cm}
\begin{thebibliography}{99}
\vspace{-0.5cm}

\bibitem{bergeretrmp} F. S. Bergeret \etal, 
Rev. Mod. Phys. \textbf{77}, 1321 (2005).

\bibitem{buzdinrmp} A. I. Buzdin, Rev. Mod. Phys. \textbf{77}, 935 (2005).

\bibitem{tanakaPRL} Y. Tanaka \etal, Phys. Rev. Lett. \textbf{99}, 037005 (2007); Y. Tanaka and A. A. Golubov, Phys. Rev. Lett. \textbf{98}, 037003 (2007).

\bibitem{EschrigLTP} M. Eschrig \etal, J. Low Temp. Phys. \textbf{147} 457 (2007).

\bibitem{Yokoyama} T. Yokoyama \etal, Phys. Rev. B \textbf{78}, 012508 (2008).

\bibitem{berezinskii} V. L. Berezinskii, 
JETP Lett. \textbf{20}, 287
(1974). 

\bibitem{volkovPRL} A. Volkov \etal, Phys. Rev. Lett. \textbf{90}, 117006 (2003)

\bibitem{linder} J. Linder \etal, Phys. Rev. B \textbf{77}, 174514 (2008).

\bibitem{kupluk} M. Yu. Kupriyanov and V. F. Lukichev, Zh. Exp. Teor. Fiz. \textbf{94}, 139 (1988) ; Yu. Nazarov, Superlatt.Microstruct. \textbf{25}, 1221 (1999). 

\bibitem{sueur_prl_08} H. le Sueur \etal, Phys. Rev. Lett. \textbf{100}, 197002 (2008).

\bibitem{Millis} T. Tokuyasu \etal, Phys. Rev. B {\bf 38}, 4504 (1988);
A. Millis \etal, Phys. Rev. B {\bf 38}, 4504 (1988). 

\bibitem{Brataas} A. Brataas \etal, Phys. Rev. Lett. {\bf 11}, 2481 (2000); A. Brataas \etal, Phys. Rep. {\bf 427}, 157 (2006).

\bibitem{hh} D.H. Hernando \etal, Phys. Rev. Lett. \textbf{88}, 047003 (2002).

\bibitem{eschrig} M. Eschrig \etal, Phys. Rev. Lett. \textbf{90}, 137003 (2003); M. Eschrig and  T. L{\"o}fwander, Nature Physics {\bf 4}, 138 (2008).

\bibitem{audrey} A. Cottet and W. Belzig, Phys. Rev. B \textbf{72}, 180503 (2005).

\bibitem{fogelstrom} M. Eschrig, Phys. Rev. B {\bf 61}, 9061 (2000);
M. Fogelstr\"om, Phys. Rev. B {\bf 62}, 11812 (2000); E. Zhao \etal, Phys. Rev. B {\bf 38}, 134510 (2004).

\bibitem{Asano} Y. Asano \etal, Phys. Rev. Lett. {\bf 98}, 107002 (2007). 

\bibitem{quasiclassical} J. W. Serene and D. Rainer, Phys. Rep. \textbf{101}, 221 (1983)

\bibitem{usadel} K. Usadel, Phys. Rev. Lett. \textbf{25}, 507 (1970).

\bibitem{yokoyama07} T. Yokoyama \etal, Phys. Rev. B \textbf{75}, 134510 (2007).

\bibitem{Braude} V. Braude and Yu. V. Nazarov, Phys. Rev. Lett. {\bf 98}, 077003 (2007).

\bibitem{meservey} R. Meservey and P. M. Tedrow, Phys. Rep. 
{\bf 238}, 173 (1994).

\bibitem{clogston} A. M. Clogston, Phys. Rev. Lett. \textbf{9}, 266 (1962);
B. S. Chandrasekhar, Appl. Phys. Lett. \textbf{1}, 7 (1962).

\end{thebibliography}
\end{document}